%% file: 2c_nica_1105_30ver9.tex
\documentclass[preprint,journal]{rmaa}

\suppressfulladdresses 
\usepackage{psfig}
\usepackage{psfrag,color}
\usepackage{booktabs}
\usepackage{longtable}
\usepackage{natbib}
\usepackage{paralist}
\usepackage{psfrag,color}
\usepackage[latin1]{inputenc}
\usepackage{afterpage,longtable,lscape}
\usepackage{rotating}
\usepackage{graphics}

\SetYear{2011} \SetConfTitle{}

\def\uv{uvby-\beta \ }

\title{$\uv$ photoelectric photometry of the open clusters NGC 6811 and NGC 6830\altaffilmark{1}}

\author{J. H. Pe\~na\altaffilmark{2},  L. Fox Machado\altaffilmark{3}, H. Garcia\altaffilmark{4}, A. Renteria \altaffilmark{2}, S. Skinner \altaffilmark{2,5}, A. Espinosa \altaffilmark{2}, E. Romero \altaffilmark{2}
\medskip }

\altaffiltext{1}{Based on observations collected at the San Pedro
M\'artir Observatory, Mexico.} \altaffiltext{2}{Instituto de Astronom\'{\i}a, Universidad Nacional Aut\'onoma de
M\'exico.}\altaffiltext{3}{Observatorio Astron\'omico Nacional, Instituto de Astronom\'{\i}a, Ensenada, M\'exico.}\altaffiltext{4}{Observatorio Astron\'omico de la Universidad Nacional Aut\'onoma de Nicaragua UNAN-MANAGUA.}\altaffiltext{5}{Universidad Nacional de Panama}

\resumen{A partir de fotometr\'{\i}a fotoel\'ectrica $\uv$ de los
c\'umulos abiertos NGC 6811 (75 estrellas) y NGC 6830 (19 estrellas) realizamos la
determinaci\'on de distancias y, por ende la pertenencia de las
estrellas a cada c\'umulo. Asimismo se determinaron la edad y el
enrojecimiento de cada uno. Dado que recientemente se han determinado estrellas variables para el primero, realizamos un estudio de dichas variables}

\abstract{From $\uv$ photometry of the open clusters NGC 6811 (75
stars), and NGC 6830 (19 stars) we were able to
determine membership of the stars to each cluster, and
fix the age and reddening for each. Since several short period stars have recently been found, we have carried out a study of these variables}


\shortauthor{Pe\~na, et al.} \shorttitle{study of NGC 6811 and NGC 6830}

\fulladdresses{ \item J.H. Pe\~na: Instituto de Astronom\'{\i}a,
UNAM, Apdo. Postal 70-264, M\'exico, D.F. M\'exico
(jhpena@astroscu.unam.mx).}

\begin{document}

\maketitle

\section{Introduction}

The study of open clusters and their short period variable stars is
fundamental in  stellar evolution. Because the cluster members
are  formed in almost the same physical conditions,  they share
similar stellar  properties such age and chemical composition. The
assumption of common age,  metallicity and distance impose strong
constraints  when modeling  an ensemble of short period pulsators
belonging to open clusters (e.g. Fox Machado et al. 2001,  2006).
Thus,  observational studies involving  variable stars in open
clusters have attracted more and more attention (e.g. Fox Machado et al.
2002, Li et al., 2002 and 2004).

A series of Papers (see  Pe\~na et al., 1994, 1998, 2003, 2007)
study the physical nature of the short period
variable stars in open clusters  by means of Str\"omgren photometry
since, once their membership to the cluster has been established,
their physical quantities can be unambiguously derived. In
particular, the determination of physical parameters of cluster
member stars from $\uv$ photometry  can be done through a comparison
with theoretical models (Lester, Gray \& Kurucz 1986, hereinafter
LGK86).

As a continuation of this study, we now present observations of the
open clusters NGC 6811 and NGC 6830.  Both clusters have no previous
published $\uv$ data.

Very recently, Luo et al. (2009) carried out a search for variable
stars in the direction of NGC 6811 with CCD photometry in B, and V bands. They detected a total of sixteen variable stars. Among
these variables, twelve were catalogued as $\delta$ Scuti stars,
while  no variability type was assigned to the remaining stars. They
claim that the twelve $\delta$ Scuti stars are all very likely
members of the cluster which makes this cluster an interesting target
for asteroseismological studies. Moreover, NGC 6811 has been selected as a asteroseismic target of the Kepler space mission (Borucki et al. 1997). Therefore, deriving accurate physical parameters for the pulsating star members is very important.

For NGC 6811 Luo et al. (2009) estimated an age of log(t)=8.76 +- 0.009 from
theoretical isochrone fitting to the color magnitude diagram (CMD hereafter)
and assuming a metallicity of Z=0.019. They determined the distance modulus and color excess of 10.59 $\pm 0.09$ and 0.12$\pm 0.05$, respectively.

To the best of our knowledge, no $\delta$  Scuti variable stars have been reported
in NGC 6830 to date.

According to the compilation of data of open clusters in Paunzen and
Mermilliod (2007, Webda), NGC 6811 has a distance [pc] of 1215;
reddening [mag] of 0.160; a distance modulus [mag] of 10.92; log age
8.799 and no metallicity reported. NGC 6830 has the following:
distance [pc] 1639; reddening [mag] 0.501; distance modulus [mag]
12.63; log age 7.572 with no metallicity determined .

\section{Observations}

These were all taken at the Observatorio Astron\'omico Nacional,
M\'exico in two different seasons, those of 2009 and of 2010. The dates are listed in Table 1.  The 1.5 m telescope to which a spectrophotometer was
attached and was utilized at all times. The first observing season was carried
out for six nights from June-July 2009. The ID charts utilized were those of WEBDA. When the NGC 6811 data  was reduced, there were several stars whose photometry showed large discrepancies with that of the literature. In view of this and due to the fact that several $\delta$ Scuti stars
were recently discovered in this cluster (Luo et al. 2009), a second observing
season was planned in 2010 to measure all of these stars in the $uvby-\beta$ system.

\begin{table*}[!ht]
\small{
\begin{center}
\caption[] {\small Log of the observing seasons.} \hspace{0.01cm}
    \label{log}
\begin{tabular}{lcccl}
\hline \hline
Epoch &   Cluster  &  Initial date  &  Final date & observers\\
      &            & year mo day    &  year mo day          \\
\hline  \hline
2009 June   & NGC 6811, NGC 6830  &  2009 06 24 & 2010 06 29 & jhp, hg, arl\\
2010 August & NGC 6811            &  2010 08 03 & 2010 08 06 & jhp, ss, er, ae\\
\hline \hline
\end{tabular}
\end{center}
jhp, J.H. Pe\~na; hg, H. Garcia; arl,  A. Renteria; ss,  S. Skinner;  er, E. Romero; ae, A. Espinosa }
\end{table*}\

\subsection{Data acquisition}
The following procedure was utilized during all these nights: each measurement consisted of at
least five ten-second integrations of each star and one ten-second
integration of the sky for the $uvby$ filters and the narrow and
wide filters that define H$\beta$. Individual uncertainties were determined by
calculating the standard deviations of the fluxes in each filter for
each star. The percentual error in each measurement is, of course, a
function of both the spectral type and the brightness of each star,
but they were observed long enough to secure sufficient photons to
get a S/N ratio of accuracy of $N/\sqrt{(N)}$ of $0.01$ mag in most
cases. Each night a series of standard stars was also observed to
transform the data into the standard system. The reduction procedure
was done with the numerical packages NABAPHOT (Arellano-Ferro \&
Parrao, 1988) which reduce the data into
a standard system, although  some were
also taken from the Astronomical Almanac (2006) for the standard bright stars. The chosen system
was that defined by the standard values of Olsen (1983) and the
transformation equations are those defined by Gr\"onbech, Olsen, \& Str\"omgren, 1976 and by Crawford \& Mander (1966). In these equations the
coefficients D, F, H and L are the slope coefficients for $(b-y)$, $
m_1$,  $c_1$ and $\beta$, respectively. The coefficients B, J and I
are the color terms  of $V$, $m_1$, and $c_1$. The averaged
transformation coefficients of each season determined from the mean of all nights are listed in Table 2
along with their standard deviations. Errors of the season were
evaluated by means of the standard stars observed. These
uncertainties were calculated through the differences in magnitude
and colors, for (V, $b-y$, $m_1$, $c_1$ and $\beta$) as (0.020,
0.017, 0.011, 0.031, 0.011) respectively, which provide a numerical
evaluation of our uncertainties. Emphasis is made on the large range
of the standard stars in the magnitude and color values: $V$:(5.4,
8.7); $(b-y)$:(0.02, 0.80); $m_1$:(0.09, 0.67); $c_1$:(0.06, 1.12)
and $\beta$:(2.53, 2.89).

The transformation equations used in the work have the following forms in which inst stands for instrumental values and std for photometric values in the standard system:

 V = A + B $(b-y)$(inst) + $y$ (inst)\\
 $(b-y)$ (std) = C + D $(b-y)$(inst)\\
 $m_1$(std) = E + F $m_1$(inst) + J $(b-y)$(inst)\\
 $c_1$(std) = G + H $c_1$(inst) + I $(b-y)$(inst) \\
 $\beta$ (std) = K + L $\beta$ (inst). \\

\noindent

Table 3 lists the photometric values of the observed stars for
the NGC 6811 cluster. In this Table we list the following: column 1, the ID number as in WEBDA, which follows Lindoff's nomenclature; columns 2 and 3 the ID from Sanders (1971) and Luo et al. (2009); the following columns, 4 to 8, the measured photometric values (N denotes the times each star was measured) the three consecutive columns list the unreddened indexes from our photometry and the final columns, the spectral type for each star, determined from the [m1] [c1] diagram and from WEBDA and Becker (1947). It is interesting to note the agreement between the spectral types deduced from the photometry with that reported by spectroscopic methods. Although there is consistency among the three values, there also is
some disagreement among them. For example, W113 has spectroscopic types of A4 and A8 from WEBDA and Becker (1947), respectively and there are some (W16 and W37) that are defined as early type stars from the spectroscopy and as later types from the photometry.  Nevertheless, we note that, in general, the spectral types are coincident within the three mentioned sources. The remaining stars we classified do not have reported spectral classes. Table 4 lists basically the same information as in Table 3, but for NGC 6830. The columns for ID of the variable stars and number of observations have been omitted since each star was measured only once.

\begin{table*}[2]
 \caption{Transformation coefficients obtained for the two observed seasons}
\begin{center}
  \begin{tabular}{lccccccc}
\hline\hline
season  &  B  &   D  &   F  &   J  &  H   &   I  &   L\\
\hline 2009 &   -0.005  & 0.975 & 1.002 & 0.037 & 1.008  & -0.067 & -1.397\\
\hline $\sigma$ &  0.051  & 0.032 & 0.045 & 0.033 & 0.049  & 0.067 & 0.031\\
\hline 2010 &   0.002  & 0.962 & 1.021 & 0.025 & 0.991  & -0.006 & -1.309\\
\hline $\sigma$ &  0.013  & 0.003 & 0.034 & 0.002 & 0.042  & 0.145 & 0.021\\
\hline\hline
\end{tabular}
\end{center}
\end{table*}

\begin{figure}
\includegraphics[height=9.5cm=width=9.5cm]{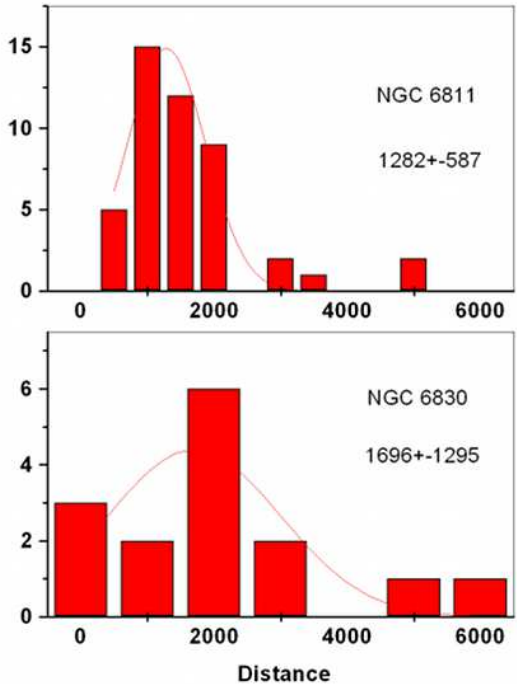}
\caption{Histogram of the distance modulus (X axis, in magnitudes)
found for the B, A and F stars in the direction of NGC 6811 upper and NGC 6830, bottom}
\end{figure}

\renewcommand{\thefootnote}{*}

\subsection{Comparison with other photometries}

Since no $\uv$ has been previously obtained for these clusters, a comparison of
our values is done with the available UBV photometry reported in WEBDA.

NGC 6811. We compared our 2009 season photometry with that reported by WEBDA. The intersection of both sets is constituted of seventy five stars, some of
them (four) showing a large
difference greater than 0.5 mag in V. There were some others (six) with differences larger than 0.1 mag. In view of this we planned and carried out a second campaign in 2010. Table 3 reports the mean values of the two observing campaigns in which N indicates the number of measurements for each star. Among the stars with large differences found in the 2009 season and WEBDA, five were measured in both observational campaigns with small differences between them. Hence, with high probability, the discrepancies cannot be attributed to our measurements because i) these stars were measured in two different seasons one year apart by different observers and ii) the measured standard stars throw reasonable values when compared to the standard literature values. Hence, the differences can be due to either a misidentification of the star by previous authors or a variable nature of these stars. Despite these differences, a linear fit between both sets yields the equation V(pp) = 0.82 + 0.97 V(WEBDA) with a correlation coefficient of 0.97 and a standard deviation of 0.25. The color relationship yields $(b-y)$ = 0.06 + 0.57 (B-V) with a correlation coefficient of 0.92 and a standard deviation of 0.09.

NGC 6830. This cluster was compared with those UBV values reported
by WEBDA in a set constituted of only seven stars. The linear fit between
both sets gave the equation V(pp) = 0.477 + 0.958 x V(WEBDA) with a
correlation coefficient of 0.997 and a standard deviation of 0.080.
The relationship in B-V and b-y gave $(b-y)$ = 0.117 + 0.573 x (B-V)
with a correlation coefficient of 0.922 and a standard deviation of
0.022.

\section{Methodology}

In order to determine the physical
characteristics of the stars in each cluster this procedure was
followed.

 The evaluation of the reddening was done by first
establishing, as was stated above, to which spectral class the stars
belonged: early (B and early A) or late (late A and F stars) types;
the later class stars (later than G) were not considered in the
analysis since no reddening determination calibration has yet been
developed for MS stars.  In order to determine the spectral type of
each star, the location of the stars in the $[m_1]-[c_1]$ diagram
was employed as a primary criteria. In Tables 3 and 4 the
photometrically determined spectral class has been indicated. The
determined spectral types compiled in the literature are also presented.

The reddening determination was obtained from the spectral types
through Str\"omgren photometry. The application of
the calibrations developed for each spectral type (Shobbrook, 1984 for O and early A
types and Nissen, 1988 for late A and F stars) were considered. No
determination of reddening was calculated for G and later spectral
types. The results of applying such calibrations are shown in Tables 5 and 6 for NGC 6811 and NGC 6830, respectively. In Table 5, the following columns are presented: columns 1 and 2 the ID (WEBDA and Luo et al., 2009) for each star; column 3, the reddening E(b-y); columns 4 to 6, the unreddened indexes $(b-y)_0$, $m_0$, and $c_0$; column 7 the H$\beta$ value, columns 8 and 9  $V_0$, and the absolute magnitude, respectively. Columns 10 and 11 show the distance modulus and the distance in parsecs. The metallicity is presented in column 12 and, finally, column 13 lists the membership to the cluster, denoted by m. The membership was determined from the Distance Modulus or distance histograms. A gaussian fit with a bin size of one was done to the bars in the histogram to all the stars and the obtained fit is presented, along with the uncertainties in each Figure. Membership then was established from the above mentioned fit. Stars within a standard deviation value from the mean were considered to be members. Those with standard deviation values slightly larger than one sigma are considered to be stars with marginal membership. In the Table, those stars that are considered to be members of the cluster are denoted by an m. Marginal membership is indicated by semi-colon, m:, those that were non-members are denoted by nm. Table 6 is analogous, but no column for variables ID is presented. Probable members are denoted by semicolon, m:

\section{Analysis}

In order to gain some insight into the clusters we must first find out which stars belong to each one. As was already mentioned, this is accomplished by constructing a histogram of the deduced distances. From the results listed in Tables 5 and 6 and shown in Figure 1, we
can establish that NGC 6811 has a distinctive accumulation of thirty-seven stars
at a distance modulus of 10.5 $\pm$ 1.0 mag, whereas NGC 6830 is merely
an association of eight early type stars at DM 11.1 $\pm$ 1.6 mag, although emphasis
should be made on the fact that we merely observed a small sample of stars in the direction of this cluster:
nineteen of the brightest stars. According to the study of Netopil et al. (2007)  four CP stars in
NGC 6830 were found. Since Str\"omgren photometry is most suitable for this topic, we checked our measured stars for the Ap determination. Unfortunately, none of our measured stars lay in the regions defined by the boxes in the $m_0$, and $c_0$ diagram where the Ap stars should be, as in Golay's (1974) Figure 124.  Hence, we cannot corroborate, nor discard the findings by Netopil et al. (2007). For NGC 6811 we determined four stars belonging to the Ap category, namely W9, W34, W105 and W491, all to the Sr-Cr-Eu class.

Age is fixed for the two determined clusters once we measured the hottest
and hence the brightest stars for each one. The
effective temperature of these hottest stars was determined by
plotting the location of all stars on the theoretical grids of
Lester, Gray and Kurucz (1986), once we evaluated the unreddened colors (Figure 2) for a
solar chemical composition. We considered this metallicity  based on the thirteen F type stars for which we determined the metallicity [Fe/H]; a mean value of -0.18 $\pm$ 0.30 was found. In the related Figures, LGK86 in the upper left corner indicates that the grids were taken from the mentioned reference of Lester, Gray and Kurucz (1986) and the specified metalicity.
We have utilized the $(b-y)$ vs. $c_0$ diagrams which allow the
determination of the temperatures with an accuracy of a few hundreds
of degrees. However, for NGC 6811, as can be seen in Figure 2, the stars are clustered together and the effective temperature cannot be easily determined. To measure their temperatures with more accuracy, a plot of $(b-y)$ vs. $\beta$ was constructed and compared with the theoretical grids of LGK86, Figure 3. The temperature for the hottest stars is around 11,700 K
for NGC 6811, whereas for NGC 6830 it is much hotter, (17,000 K). Once
membership has been established, age is determined after
calculating the effective temperature through the calibrations of
Meynet, Mermilliod and Maeder (1993) for open clusters; a log age of
8.266 (1.845 x $10^{8}$yr) is found from the relation -3.611 log $\log T_{\rm eff}$ +
22.956 valid in the range log $\log T_{\rm eff}$ within the limits [3.98, 4.25] for
NGC 6811; whereas for NGC 6830 the relation log(age) = -3.499 log $\log T_{\rm eff}$
+ 22.476 valid in the range [4.25, 4.56] yields log (age) of 7.69
(4.89 x $10^{7}$ yr).

These determinations are confirmed by constructing the color-magnitude diagram of NGC 6811 and NGC 6830 which are shown in Figs 4 and 5, respectively.
The unreddened magnitudes  [$(b-y)_{0},M_{V}$]  of cluster members taken from
Tables 5 (NGC 6811) and 6 (NGC 6830) are shown with filled circles.
In each plot  two theoretical isochrones  in the Str\"omgren photometric system
are shown with solid and dashed lines. The metallicity and  ages are indicated in the Figures.
The theoretical isochrones were obtained from the
Padova database (Girardi, Bertelli, Bressan, et al. 2003). As can be seen, the isochrones match the observed color-magnitude diagram with the ages and distance derived in the present paper.

\begin{figure}
\includegraphics[width=\columnwidth]{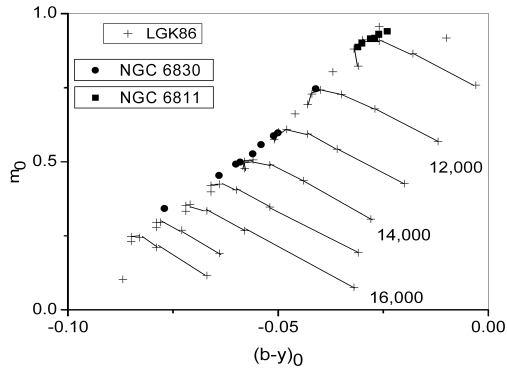}
\caption{Location of the unreddened points of the two clusters in the LGK86
grids. Squares, those of NGC 6811; dots, of NGC 6830.}
\end{figure}

\begin{figure}
\includegraphics[height=6.5cm=width=6.5cm]{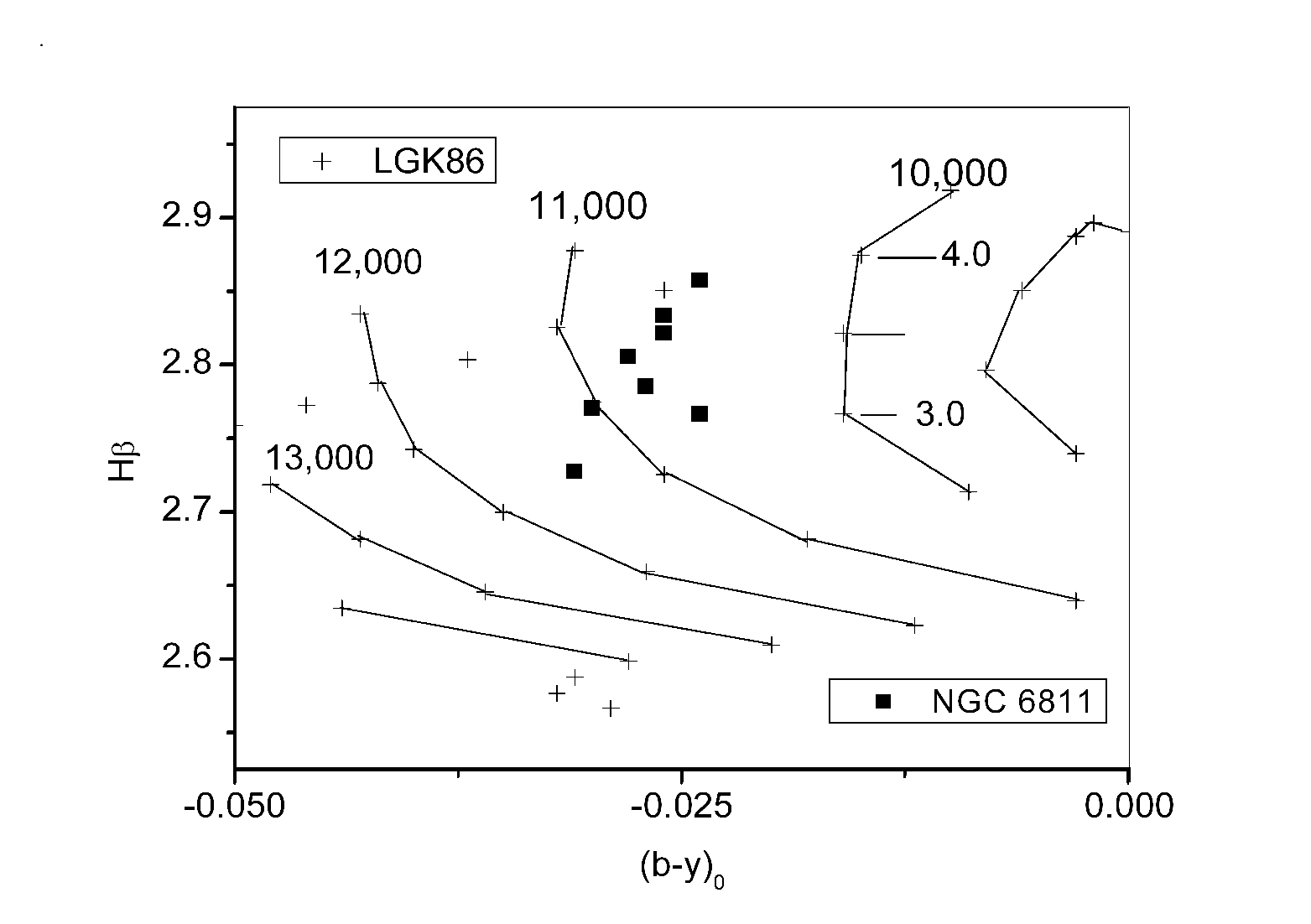}
\caption{Location of the unreddened points of the hot stars of NGC 6811 cluster in the LGK86
grids.}
\end{figure}

\begin{figure}
\includegraphics[height=7.0cm=width=7.0cm]{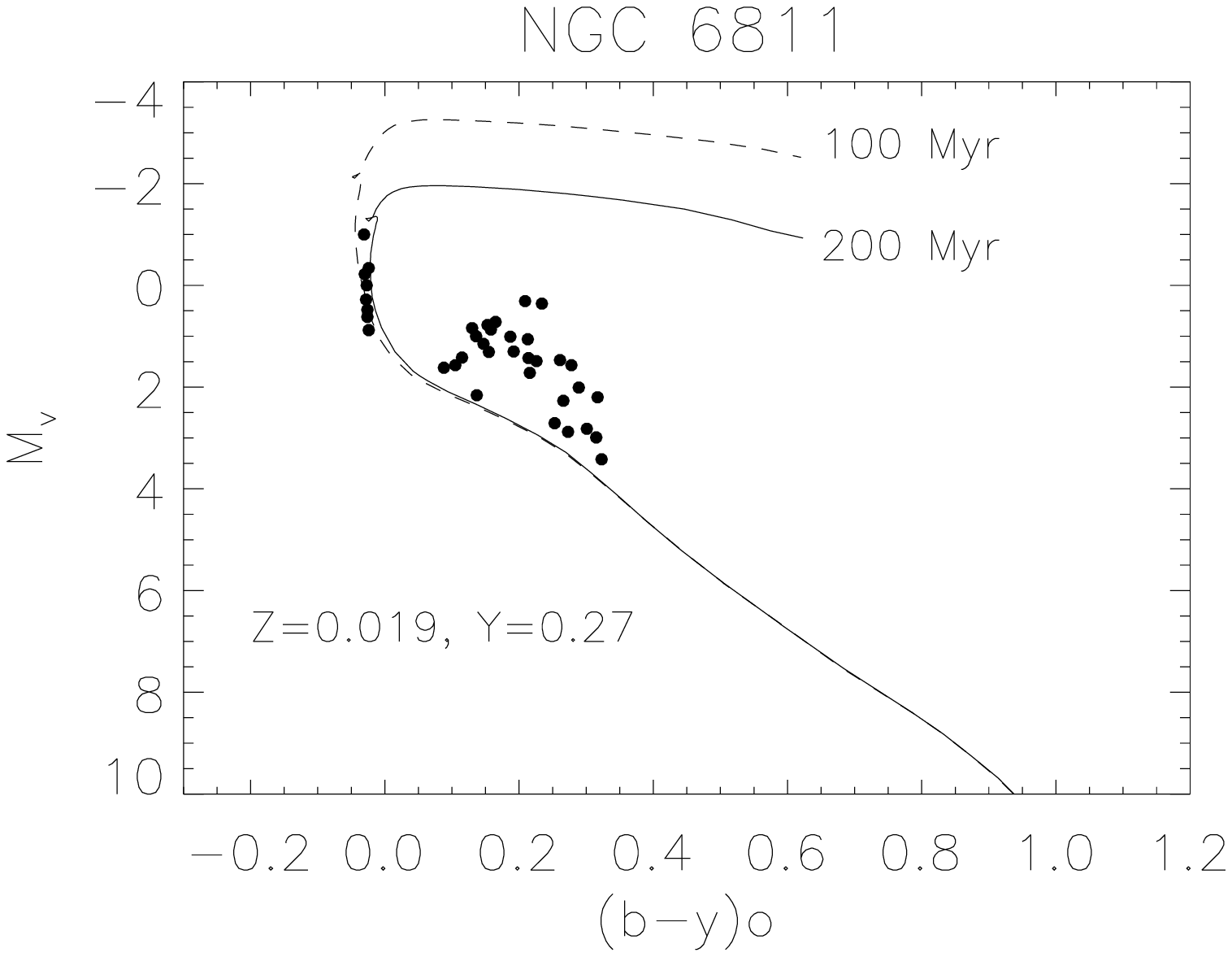}
\caption{Color-magnitude diagram of the NGC 6811 cluster considering only the cluster members. The target stars are represented by filled circles. Theoretical isochrones of
100 Myr (dashed line) and 200 Myr (solid line) computed with Z=0.019 are shown.}
\end{figure}

\begin{figure}
\includegraphics[height=7.0cm=width=7.0cm]{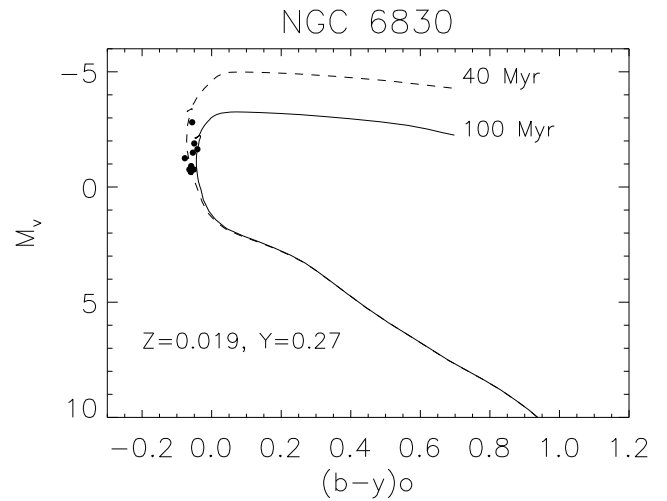}
\caption{Color-magnitude diagram of the NGC 6830 cluster considering only the cluster members. The target stars are represented by filled circles. Theoretical isochrones of
40 Myr (dashed line) and 100 Myr (solid line) computed with Z=0.019 are shown.}
\end{figure}

\section{Variable stars in NGC 6811}
As was stated in the Introduction, Luo et al. (2009) performed time-series photometric observations of the open cluster NGC 6811 to search for variable stars. These observations were carried out during five nights from June 6 to July 24, 2008 utilizing the 85 cm telescope of the Xonglong Station of the National Astronomical Observatories of the Chinese Academy of Sciences. The instrumentation they used was 1024 x 1024 CCD camera with a field of
view of 16.5'x16.5' with standard Johnson-Cousin-Bessell filters in B and V bands, of which they obtained 750 CCD frames in each band. Sixteen certain variable
stars were detected or confirmed from that survey, namely V1-V7 and V10-V18
following the variable name list of Van Cauteren et al. (2005) (see Table 1 by Luo et al. 2009). Among these, twelve stars were catalogued as Delta Scuti variables based upon the light curves (V1-V7 and V10-V14).
The omitted variables, V8 and V9, were outside of their field-of-view. In particular, Luo et al. (2009) discovered variability in V10-V18;  four of
them (V10, V12, V15, V16)  had been just reported  as suspected variables by
Rose and Hintz (2007), while the variability of V1-V7 was discovered by Van
Cauteren et al. (2005). On the other hand, nine stars reported as variables by Rose and Hintz (2007) were not confirmed by Luo et al. (2009). One explanation provided for this inconsistency was that the amplitude of light variations
was too low to be detected. Luo et al. 2009 also determined the membership probabilities of twelve variables (V1-V5 and V10-16) through
the proper-motion membership probabilities (PMP)
listed by Sanders (1971). From these values they claim that with high probability
all of the twelve stars (V1-V3, V5, V10-V16) are cluster members, except for V4. For the the stars without PMP data, namely V6, V7, V17 and V18, based
on their position in the CMD diagram, they concluded that
the first two  are most likely  members of the cluster whereas
the last two are probably field stars.

On the night of  August 6, 2010 (UT) we carried out very short span of observations in differential photometric mode. The variables we considered were chosen due to their nearness and were, in the notation of Luo et al. (2009): V2, V4, V11 and V14 with W5 and W99 as reference and check stars. Although the time span we observed was too short to detect long period variation, the only star which showed clear variation was V4, with two clearly discernible peaks and of relatively large amplitude of variation 0.188 mag and a period of 0.025 d.

From our cluster membership determinations on a star-to-star basis, the conclusion we reach is slightly different from the previous assertions regarding variability. Memberships are determined for V1, V3, V4, V5, V10, V11, V13 and V16. Marginal membership for V12 and V14, non-membership for V15 and we were unable to determine membership  for the remaining stars mainly because they do not belong to the spectral classes B, A or F and belong to a latter spectral type which makes them unlikely to be $\delta$ Scuti type variables. From the location of these variables in the theoretical grids of LGK86, Figures 6 and 7, we determine their temperatures.

\begin{figure}
\includegraphics[height=7.0cm=width=7.0cm]{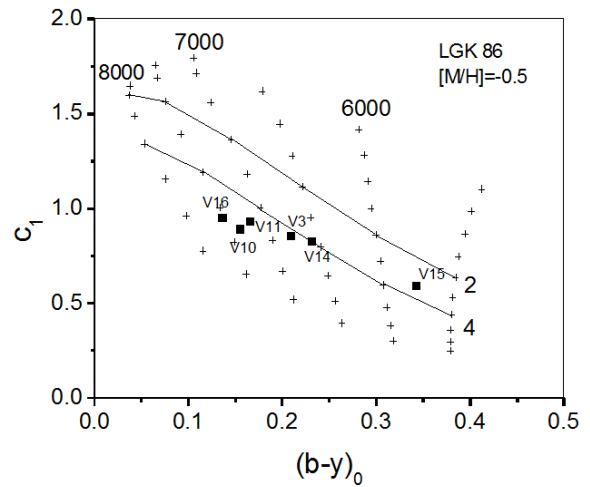}
\caption{Location of the $\delta$ Scuti stars of NGC 6811 in the theoretical grids of LGK86.}
\end{figure}

\begin{figure}
\includegraphics[height=7.0cm=width=7.0cm]{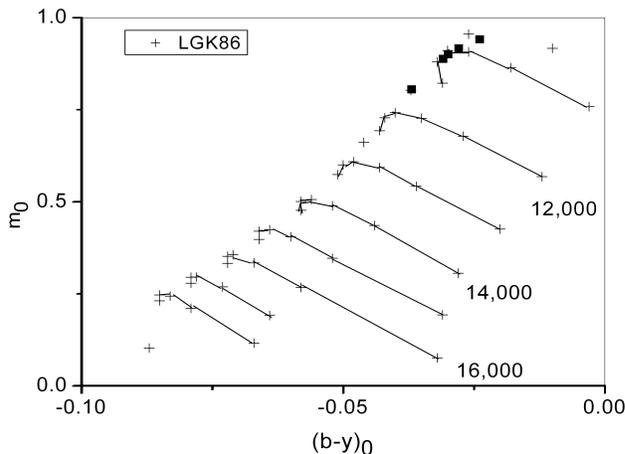}
\caption{Location of the hot variable stars of NGC 6811 in the theoretical grids of LGK86.}
\end{figure}

\section{Confidence of the results}
As has been said in previous sections, the high accuracy of each observed star
was attained by multiply observing each star in sequences of five 10
sec integrations. Hence, mean values and standard deviations were
calculated to determine the signal/noise ratio. In all
cases enough star counts were secured to attain a signal to noise
ratio large enough to determine an accuracy better than $0.01$ mag.
Nevertheless, it is obvious that the brighter stars were more
accurately observed than the fainter ones. Quoting Nissen (1988) ``as
expected from photon statistics considerations the average mean
errors increase as we go to fainter magnitudes''. Unfortunately,
since the aim of this project was to observe as many stars as
possible, most of them were observed only twice, and a few, only
once. The uncertainties of the season were determined from the
differences between the derived magnitude of the standard stars vs. reported values in the literature. The average values of such
differences are $\Delta$ $(V, b-y, m_1, c_1)$ = (0.008, 0.005, -0.004,
0.012); on most nights at least ten standard stars were observed
but this figure increased to 15 on some nights. The number of the
whole sample of standards data points, due to the large time span of
the season, was considerable, adding up to 80 points of standard
stars.

To calculate the propagation of errors for the reddening (in Nissen's (1988)
work, section 3), the intrinsic color index $(b-y)_0$ has served to
determine the individual color excess, $E(b-y) = (b-y) - (b-y)_0$
and, as in his paper, assuming the photometric mean errors given for
our observations, although larger than the work by Nissen (1988), we
do expect a mean error $E(b-y)$ of close to that derived by Nissen
of $0.011$ for F stars and of $0.009$ for A stars since our errors
are not exceedingly different.

\setcounter{table}{3} 
\begin{table*}[!t]
 \caption{$\uv$ photoelectric photometry of the open cluster NGC 6830 }
 \begin{center}
  \begin{tabular}{rrlrllllll}
\hline \hline

ID&V&$(b-y)$ &$ m_1$&$c_1$&$\beta$&   SpTyp& &sptp\\
\hline \hline
&&&&&&Phe&Spc\\
\hline \hline

5&9.849&0.266&0.013&0.554&2.721& B6V & B7 V & B5 III \\
7&11.176&0.353&0.007&0.822&2.689& B8V & A0 V & B7 III \\
8&11.540&0.373&-0.018&0.716&2.616& B7V & B7 IV & \\
49&10.956&0.456&0.103&0.875&2.751& AV5 &  &\\
4&12.623&0.600&-0.100&0.471&2.685& B3V  & & \\
2258&13.003&0.546&0.086&0.484&2.721& F9V  &  &\\
2257&12.166&0.390&-0.008&0.584&2.706& B7V & & \\
26&12.309&0.564&0.160&0.255&2.661& G1V &  &\\
164&12.240&0.460&0.093&0.819&2.766& A8V &  &\\
25&12.465&0.333&0.012&0.661&2.718& B8V &  &\\
24&11.644&0.391&-0.034&0.612&2.635& B5V  & B6 V &  B6 IV \\
2&10.625&0.329&-0.017&0.606&2.603& B5V &  &\\
3&12.888&1.620&0.383&0.256&2.569& K G  & & \\
39&11.192&0.386&-0.051&0.681&2.668& B8I & B5 IV & \\
2275&11.909&1.160&0.306&-0.047&2.576& B7V &  &\\
13&12.787&0.767&0.134&0.380&2.605& K0V  &  &\\
14&11.995&0.348&0.017&0.532&2.714& B6V  & B6 V & B6 IV \\
15&12.659&0.407&-0.035&0.646&2.681& B6V &  &\\
16&12.456&0.408&0.108&0.807&2.794& K &  &\\

\hline \hline
\hline \hline

\end{tabular}
\end{center}
\end{table*}

\setcounter{table}{5} 
\begin{table*}[!t]
 \caption{Reddening, unreddened parameters and distance of the open cluster NGC 6830}
 \begin{center}
  \begin{tabular}{rccccccrrrrrc}
  \hline \hline

ID   &  $E(b-y)$&$(b-y)_0$ &$ m_0$&$c_0$&$\beta$&$V_0$&$M_V$&DM&DST&[Fe/H]&Mbr\\
\hline
26   & 0.244 &  0.320 & 0.233 & 0.206 & 2.661 & 11.26 &  5.67 &  5.59 &  131 & 0.85 & nm\\
2258 & 0.310 &  0.236 & 0.179 & 0.422 & 2.721 & 11.67 &  4.10 &  7.57 &  327 &  & nm\\
13   & 0.371 &  0.396 & 0.245 & 0.306 & 2.605 & 11.19 &  3.43 &  7.76 &  357 & 0.46 & nm\\
164  & 0.291 &  0.169 & 0.180 & 0.761 & 2.766 & 10.99 &  1.90 &  9.09 &  658 &  & nm\\
5    & 0.326 & -0.060 & 0.111 & 0.492 & 2.721 &  8.45 & -0.65 &  9.10 &  661 &  & m:\\
4    & 0.677 & -0.077 & 0.103 & 0.342 & 2.685 &  9.71 & -1.25 & 10.96 & 1559 &  & m\\
14   & 0.412 & -0.064 & 0.141 & 0.454 & 2.714 & 10.22 & -0.76 & 10.98 & 1571 &  & m\\
7    & 0.394 & -0.041 & 0.125 & 0.747 & 2.689 &  9.48 & -1.64 & 11.12 & 1676 &  & m\\
2257 & 0.449 & -0.059 & 0.127 & 0.499 & 2.706 & 10.23 & -0.91 & 11.14 & 1692 &  & m\\
39   & 0.436 & -0.050 & 0.080 & 0.598 & 2.668 &  9.32 & -1.89 & 11.21 & 1745 &  & m\\
25   & 0.384 & -0.051 & 0.127 & 0.588 & 2.718 & 10.81 & -0.76 & 11.57 & 2063 &  & m\\
15   & 0.461 & -0.054 & 0.103 & 0.558 & 2.681 & 10.68 & -1.49 & 12.16 & 2709 &  & m\\
24   & 0.447 & -0.056 & 0.100 & 0.527 & 2.635 &  9.72 & -2.81 & 12.53 & 3200 &  & m\\
2    & 0.385 & -0.056 & 0.098 & 0.533 & 2.603 &  8.97 & -4.40 & 13.37 & 4719 &  & nm\\
8    & 0.420 & -0.047 & 0.108 & 0.636 & 2.616 &  9.73 & -4.08 & 13.81 & 5785 &  & nm\\

\hline \hline
\end{tabular}
\end{center}
\end{table*}

\section{Discussion}

New $\uv$ photoelectric photometry has been acquired and is
presented for the brightest stars in the direction of two open
clusters NGC 6811 and NGC 6830. From the observed stars in
the field, some were determined to be early type stars, either B or
A. Using the calibrations to determine reddening and distance for
these stars, distances for the clusters have been determined.
Unreddened indexes in the LGK86 grids allowed us to determine the
effective temperature of the hottest stars and hence, the age of the cluster.

A brief discussion of each cluster is presented. Table 7 lists the
previous  knowledge and the newly determined characteristics of the clusters.

NGC 6811. Considering the classical UBV photometry compiled for this
cluster, very little can be deduced about its properties. No clear
distinction in the color-color diagram B-V vs U-B can be drawn; the
same conclusion is reached from its HR diagram. From our results we
have determined 37 stars belong to a cluster. Since they are
the brightest, the conclusions on the age which agrees with that
previously determined is also correct. We have found that the
cluster is farther, its extinction is less and it is younger than previously assumed. The goodness of our method has been previously tested, as in the case of the open cluster Alpha Per
(Pe\~na \& Sareyan, 2006) against several sources which consider
proper motion studies as well as results from Hipparcos and Tycho
data basis. Hence, we feel that our results throw new light
regarding membership to this cluster.

There have been several previous works in which membership probabilities were considered. Table 8 lists the identification numbers from several studies, namely those of Becker (1947), Sanders (1971), Lindoff (1972), Barkhatova (1978) and the more recent ones, the compilation by Kharchenko et al. (2005) and the variability study of Luo et al. (2009). We have repeated part of the information on  the distance provided in Table 8 in order to support the conclusions based on the last columns of the Table which present the membership probability obtained in the present paper (PP), that of Sanders (1971) and those reported by the compilation of Kharchenko et al. (2005) based in the studies of proper motion, photometry and position of the stars.
Membership probabilities, if compared with those of Sanders' (1971), are in rough agreement: all but two stars (W92 and W146) that we assign cluster membership are not assigned as members according to Sanders' probabilities, but it is equally true that those which we define as non-members are determined to be members by Sanders (1971). When the comparison is done with those probabilities of the compilation of Kharchenko et al. (2005) the conclusions are equally in agreement. There are two stars, W18 and W45 we define as members that Kharchenko et al. (2005) find to be non-members from the proper motion studies but members with the other two criteria. In conclusion the comparison of our results with the others support our findings particularly because the results obtained from $\uv$ photometry are more accurate.

The DM and reddening determined from our photometry, although discordant from those derived from UBV photometry, is in agreement with that of Glushkhova et al. (1999) which, from radial velocity measurements for 60 late-type stars and UBVRI photoelectric photometry refined the distance modulus to the cluster to be 10.47 $\pm 0.08$ mag and E(B-V) of 0.12 $\pm 0.02$, in agreement with the values we derived

We were able to determine membership in the NGC 6811 open cluster of several variable stars. We found that six stars V1, V4, V10, V11, V13 and V16 are
cluster members, on the contrary V12, V14 and V15 are definitively
non-member stars. For the rest not much can be said. Accurate temperature determination was done for each one.

NGC 6830. Again, since no previous $\uv$ exists, knowledge of the
cluster rests on UBV photometry. Both the color-color and the
color-magnitude diagrams do not show a clear main sequence which
make the distance, age and reddening determinations ambiguous. We
only measured nineteen stars, but with this
small sample we determined some clustering of stars. Our findings coincide, within the uncertainties, with the previous distance, reddening and age determinations. Of course, much more data is needed
to unambiguously establish the true nature of this cluster, but we emphasize that, since we observed all the bright stars, our conclusions are correct.

\setcounter{table}{6} 
\begin{table*}[!t]
 \caption{Compiled characteristics for NGC 6830 and NGC6811}
 \begin{center}
  \begin{tabular}{llcccc}
\hline\hline
Cluster & Source & Log age  & Reddening $[mag]$E(B-V) & Distance [kpc] & Metallicity  \\
 \hline
NGC 6830 & Barkhatova (1957)             &      &       & 1.68 &      \\
         & Hoag \& Applequist (1965)     &      & 0.51  & 1.38 &      \\
         & Becker \& Fenkart (1971)      &      & 0.58  & 1.47 &      \\
         & Moffat (1972)                 & 8.0  & 0.56  & 1.70 &      \\
         & Glushkova et al. (1999)       &      & 0.12  & 1.24 &      \\
         & Dias et al. (2002)            & 7.57 & 0.50  & 1.64 &      \\
         & Kharchenko et al. (2005)      & 7.52 & 0.50  & 1.64 &      \\
         & Paunzen \& Mermilliod (Webda) & 7.57 & 0.50  & 1.64 &      \\
         & PP                            & 7.69 & 0.63  & 1.88 & +0.13\\
NGC 6811 & Luo et al. (2009)             & 8.76 & 0.12  & 1.31 & +0.02\\
         & Paunzen \& Mermilliod (Webda) & 8.80 & 0.16  & 1.22 &      \\
         & PP                            & 8.27 & 0.14  & 1.64 & -0.02\\
\hline \hline
\end{tabular}
\end{center}
\end{table*}

\acknowledgments

We would like to thank the staff of the OAN and E. Romero for their assistance in
securing the observations. This work was partially supported by
PAPIIT IN110102 and IN114309. HG, ER, SS, AE and ARL thank the IA-UNAM for the opportunity to carry out the observations. JHP thanks the hospitality of the UNAN. Typing and proofreading were done by J. Orta, and J. Miller, respectively. C. Guzm\'an, F. Salas and A. Diaz assisted
us in the computing. This research has made use of the Simbad
databases operated at CDS, Strasbourg, France and NASA ADS Astronomy
Query Form.

\clearpage\onecolumn

\begin{landscape}
  \input Tbl_3.tex

  \clearpage
  \input Tbl5_unrddnd.tex

    \clearpage
  \input Tbl8.tex

\end{landscape}

\end{document}

%% file: Tbl_3.tex
\setcounter{table}{2}
\tablecols{15}
\tabcaption{$\uv$ photoelectric photometry of the open
cluster NGC 6811}

\def\ColumnHeaders{
WBD & V & $(b-y)$ & $m_1$ & $c_1$ & bt & sV & sby & sm$_1$ & sc$_1$ & sbt & N &\multicolumn{2}{c}{ Spectral Type}\\
    &   &         &       &       &    &    &     &        &        &     &   & Photometry & WBD}
\begin{longtable}{rcccccrllllllll}
  \toprule
  \ColumnHeaders\\
  \midrule
  \endfirsthead

  \tabcaptioncontinued
  \toprule
  \ColumnHeaders\\ \midrule
  \endhead

  \bottomrule
  \endfoot

  4 & 12.681 & 0.259 & 0.127 & 0.861 & 2.706 &  &  &  &  &  & 1 & A3V & A7\\
  5 & 11.795 & 0.168 & 0.185 & 0.950 & 2.770 & 0.067 & 0.028 & 0.015 & 0.029 & 0.062 & 28 & A5V & A2\\
  6 & 14.463 & 0.786 & 0.291 & 0.554 & 2.538 & 0.076 & 0.111 & 0.108 & 0.226 & 0.071 & 3 & $>$G &\\
  7 & 14.381 & 0.399 & 0.125 & 0.388 & 2.588 & 0.201 & 0.071 & 0.221 & 0.104 & 0.020 & 3 & F7V &\\
  8 & 14.110 & 0.356 & 0.142 & 0.498 & 2.630 & 0.150 & 0.053 & 0.153 & 0.115 & 0.055 & 3 & F7V &\\
  9 & 12.081 & 0.166 & 0.218 & 0.959 & 2.790 & 0.080 & 0.015 & 0.029 & 0.067 & 0.077 & 3 & Ap & A1\\
 10 & 14.234 & 0.339 & 0.223 & 0.346 & 2.525 & 0.283 & 0.073 & 0.065 & 0.153 & 0.180 & 3 & G2V &\\
 12 & 14.363 & 0.455 & 0.148 & 0.387 & 2.540 & 0.098 & 0.067 & 0.057 & 0.142 & 0.094 & 3 & G0V &\\
 13 & 15.055 & 0.567 & 0.058 & 0.542 & 2.585 & 0.264 & 0.158 & 0.164 & 0.277 & 0.037 & 2 & F7V &\\
 14 & 13.672 & 0.765 & 0.399 & 0.174 & 2.510 & 0.182 & 0.029 & 0.041 & 0.022 & 0.040 & 2 & $>$G &\\
 16 & 12.196 & 0.235 & 0.165 & 0.927 & 2.781 &       &       &       &       &       & 1 & A5V & A4\\
 18 & 12.118 & 0.236 & 0.144 & 0.966 & 2.806 &       &       &       &       &       & 1 & A5V & A4\\
 22 & 13.034 & 0.522 & 0.267 & 0.230 & 2.529 & 0.019 & 0.018 & 0.005 & 0.023 & 0.039 & 2 & K0V &\\
 23 & 13.981 & 0.558 & 0.227 & 0.185 & 2.655 &       &       &       &       &       & 1 & $>$G &\\
 24 & 11.245 & 0.613 & 0.327 & 0.318 & 2.564 & 0.044 & 0.020 & 0.012 & 0.052 & 0.019 & 2 & $>$G & G8\\
 26 & 11.414 & 0.197 & 0.186 & 0.985 & 2.795 &       &       &       &       &       & 1 & A5V &\\
 31 & 13.308 & 0.327 & 0.131 & 0.635 & 2.760 &       &       &       &       &       & 1 & F5V &\\
 32 & 11.351 & 0.640 & 0.349 & 0.322 & 2.583 &       &       &       &       &       & 1 & $>$G & G\\
 33 & 11.917 & 0.232 & 0.149 & 0.950 & 2.771 &       &       &       &       &       & 1 & A5V & A2\\
 34 & 11.623 & 0.204 & 0.204 & 0.956 & 2.829 &       &       &       &       &       & 1 & A5p & B9\\
 35 & 13.859 & 0.325 & 0.206 & 0.406 & 2.559 & 0.084 & 0.071 & 0.052 & 0.071 & 0.020 & 2 & G2V &\\
 36 & 13.221 & 0.283 & 0.180 & 0.561 & 2.614 & 0.065 & 0.062 & 0.049 & 0.009 & 0.052 & 2 & G0V &\\
 37 & 11.113 & 0.182 & 0.181 & 0.933 & 2.766 & 0.034 & 0.013 & 0.006 & 0.023 & 0.040 & 29 & A5V-F5Ib\\
 38 & 13.206 & 0.662 & 0.382 & 0.330 & 2.535 & 0.052 & 0.079 & 0.013 & 0.066 & 0.128 & 2 & $>$G &\\
 39 & 11.528 & 0.212 & 0.157 & 0.961 & 2.728 & 0.069 & 0.020 & 0.012 & 0.042 &       & 29 & A5V & A4\\
 40 & 13.070 & 0.111 & 0.147 & 0.632 & 2.673 & 0.127 & 0.062 & 0.073 & 0.100 & 0.071 & 2  & A1V &\\
 41 & 12.014 & 0.148 & 0.195 & 0.956 & 2.741 & 0.100 & 0.035 & 0.017 & 0.037 &       & 29 & A5V &\\
 42 & 12.568 & 0.190 & 0.184 & 0.829 & 2.698 & 0.172 & 0.055 & 0.028 & 0.049 & 0.059 & 29 & A8V &\\
 43 & 12.743 & 0.268 & 0.176 & 0.785 & 2.658 & 0.023 & 0.000 & 0.001 & 0.049 & 0.030 & 2  & A8V &\\
 44 & 12.046 & 0.169 & 0.182 & 0.935 & 2.757 & 0.082 & 0.033 & 0.016 & 0.032 & 0.029 & 29 & A5V &\\
 45 & 12.705 & 0.204 & 0.194 & 0.831 & 2.735 & 0.142 & 0.027 & 0.015 & 0.040 & 0.039 & 3 & A8V & A5\\
 46 & 12.958 & 0.329 & 0.116 & 0.533 & 2.645 & 0.058 & 0.006 & 0.017 & 0.047 & 0.043 & 3 & F5V &\\
 47 & 13.649 & 0.383 & 0.099 & 0.389 & 2.613 & 0.136 & 0.020 & 0.046 & 0.065 & 0.083 & 3 & F5V &\\
 49 & 12.422 & 0.218 & 0.142 & 0.978 & 2.822 &       &       &       &       &       & 1 & A0V & A5\\
 51 & 13.418 & 0.269 & 0.139 & 0.592 & 2.590 & 0.076 & 0.094 & 0.115 & 0.125 & 0.035 & 2 & F5V &\\
 53 & 12.754 & 0.178 & 0.191 & 0.898 & 2.774 & 0.076 & 0.020 & 0.030 & 0.045 & 0.085 & 3 & A5V &\\
 54 & 12.356 & 0.266 & 0.136 & 0.757 & 2.713 & 0.027 & 0.025 & 0.029 & 0.027 & 0.078 & 3 & A8V &\\
 56 & 12.160 & 0.247 & 0.149 & 0.787 & 2.691 & 0.013 & 0.025 & 0.024 & 0.055 & 0.037 & 2 & A8V &\\
 57 & 14.039 & 0.439 & 0.161 & 0.357 & 2.670 & 0.149 & 0.196 & 0.161 & 0.114 & 0.073 & 4 & G0V &\\
 58 & 14.350 & 0.622 & 0.219 & 0.151 & 2.650 & 0.187 & 0.231 & 0.122 & 0.256 & 0.132 & 3 & K0V &\\
 62 & 12.846 & 0.224 & 0.163 & 0.728 & 2.714 &       &       &       &       &       & 2 & A8V &\\
 63 & 15.234 & 0.979 & 0.846 & 0.714 &       & 0.193 & 0.569 & 0.401 & 0.677 & 0.294 & 2 & $>$G &\\
 64 & 13.609 & 0.352 & 0.088 & 0.557 & 2.897 & 0.041 & 0.016 & 0.056 & 0.021 & 0.366 & 2 & F2V &\\
 65 & 14.450 & 0.460 & 0.199 & 0.217 & 2.510 & 0.087 & 0.164 & 0.106 & 0.057 & 0.103 & 2 & G2V &\\
 68 & 10.850 & 0.272 & 0.152 & 0.869 & 2.737 &       &       &       &       &       & 1 & A8V & A2\\
 70 & 10.925 & 0.298 & 0.139 & 0.876 & 2.713 &       &       &       &       &       & 1 & A5V & A4\\
 71 & 13.716 & 0.388 & 0.154 & 0.473 & 2.551 &       &       &       &       &       & 1 & G0V & A2\\
 73 &  9.851 & 0.992 & 0.830 & 0.166 & 2.551 &       &       &       &       &       & 1 & $>$G & K5\\
 74 & 12.322 & 0.415 & 0.171 & 0.305 & 2.568 & 0.078 & 0.007 & 0.003 & 0.028 & 0.008 & 2 & G0V\\
 77 & 13.860 & 0.257 & 0.206 & 0.461 & 2.650 & 0.120 & 0.040 & 0.041 & 0.028 & 0.025 & 2 & G0V\\
 78 & 13.664 & 0.270 & 0.167 & 0.567 & 2.676 & 0.070 & 0.023 & 0.011 & 0.040 & 0.037 & 2 & F7V\\
 79 & 10.393 & 0.935 & 0.719 & 0.146 & 2.519 &       &       &       &       &       & 1 & $>$G \\
 82 & 13.025 & 0.705 & 0.341 & 0.333 & 2.546 &       &       &       &       &       & 1 & $>$G \\
 85 & 12.860 & 0.308 & 0.134 & 0.545 & 2.684 &       &       &       &       &       & 1 & F5V\\
 86 & 13.389 & 0.334 & 0.106 & 0.490 & 2.662 &       &       &       &       &       & 1 & F2V\\
 87 & 12.622 & 0.380 & 0.206 & 0.399 & 2.617 &       &       &       &       &       & 1 & G2V\\
 92 & 12.260 & 0.221 & 0.166 & 0.722 & 2.704 &       &       &       &       &       & 1 & A8V\\
 99 & 11.962 & 0.168 & 0.184 & 0.957 & 2.817 & 0.036 & 0.036 & 0.028 & 0.028 & 0.078 & 2 & A5V & B9\\
101 & 10.682 & 0.671 & 0.393 & 0.333 & 2.569 &       &       &  &  &  & 1 & $>$G &\\
102 & 12.882 & 0.586 & 0.221 & 0.287 & 2.562 &       &       &  &  &  & 1 & K0V &\\
105 & 12.419 & 0.230 & 0.241 & 0.850 & 2.802 &       &       & &  &  & 1 & Ap & A7\\
106 & 11.379 & 0.301 & 0.138 & 0.796 & 2.713 &       &       &  &  &  & 1 & A8V & A3\\
107 & 12.726 & 0.419 & 0.123 & 0.414 & 2.640 &       &  &  &  &  & 1 & F7V &\\
112 & 12.825 & 0.372 & 0.146 & 0.412 & 2.663 &       &  &  &  &  & 1 & F9V &\\
113 & 11.471 & 0.233 & 0.144 & 0.990 & 2.767 &       &  &  &  &  & 1 & A5V & A4\\
114 & 12.145 & 0.229 & 0.141 & 0.967 & 2.786 & 0.020 & 0.033 & 0.019 & 0.053 & 0.035 & 3 & A5V & B7\\
115 & 11.551 & 0.220 & 0.169 & 0.787 & 2.759 & 0.135 & 0.040 & 0.023 & 0.076 & 0.025 & 3 & A3V & A1\\
118 & 12.352 & 0.469 & 0.174 & 0.399 & 2.548 &       &  &  &  &  & 1 & G2V &\\
122 & 12.825 & 0.372 & 0.146 & 0.412 & 2.663 &       &  &  &  &  & 1 & F9V &\\
123 & 13.970 & 0.648 & 0.689 & 0.386 & 2.579 &       & &  &  &  & 1 & $>$G &\\
133 & 12.069 & 0.576 & 0.346 & 0.230 & 2.510 &       &  &  &  &  & 1 & $>$G & G2\\
139 & 13.197 & 0.358 & 0.082 & 0.569 & 2.646 &  &  &  &  &  & 1 & F2V &\\
146 & 12.504 & 0.353 & 0.151 & 0.475 & 2.650 &  &  &  &  &  & 1 & F9V &\\
147 & 12.129 & 0.168 & 0.200 & 0.972 & 2.847 &  &  &  &  &  & 1 & A5V & A4\\
178 &  9.909 & 1.039 & 0.839 & 0.225 & 2.835 &  &  &  &  &  & 1 & $>$G &\\
218 & 12.087 & 0.194 & 0.157 & 0.982 & 2.858 &  &  &  &  &  & 1 & A5V & A5\\
489 & 11.000 & 0.252 & 0.179 & 0.855 & 2.748 &  & &  &  &  & 1 & A5V &\\
491 & 13.681 & 0.240 & 0.241 & 0.857 & 2.889 &  &  &  &  &  & 1 & Ap &\\
V17 & 15.009 & 0.560 & 0.067 & 0.157 & 2.468 &  &  &  &  &  & 1 & F9V &
\end{longtable}

%% file: Tbl5_unrddnd.tex
\setcounter{table}{4}
\tablecols{13}
\tabcaption{Reddening, unreddened parameters and distance of the open cluster NGC 6811}
\def\ColumnHeaders{
 ID & $E(b-y)$ & $(b-y)_0$ & $ m_0$ & $c_0$ & $\beta$ & $V_0$ & $M_V$ & DM & DST & $[Fe/H]$ & Memb
}
\begin{longtable}{rcccccccccccc}
  \toprule
  \ColumnHeaders\\ \midrule
  \endfirsthead

  \tabcaptioncontinued
  \toprule
  \ColumnHeaders\\ \midrule
  \endhead

  \bottomrule
  \endfoot
112  & 0.082 & 0.290 & 0.171 & 0.396 & 2.663 & 12.47 & 3.82 &  8.7 &  537 & 0.1 & NM\\
122  & 0.082 & 0.290 & 0.171 & 0.396 & 2.663 & 12.47 & 3.82 &  8.7 &  537 & 0.1 & NM\\
107  & 0.108 & 0.311 & 0.155 & 0.392 & 2.640 & 12.26 & 3.47 &  8.8 &  574 &-0.2 & NM\\
 31  & 0.134 & 0.193 & 0.171 & 0.608 & 2.760 & 12.73 & 3.47 &  9.3 &  713 &     & M\\
489  & 0.072 & 0.180 & 0.201 & 0.841 & 2.748 & 10.69 & 1.32 &  9.4 &  749 &     & M\\
146  & 0.052 & 0.301 & 0.167 & 0.465 & 2.650 & 12.28 & 2.82 &  9.5 &  780 & 0.0 & M\\
 68  & 0.085 & 0.187 & 0.177 & 0.852 & 2.737 & 10.48 & 1.01 &  9.5 &  786 &     & M\\
 34  & 0.099 & 0.105 & 0.234 & 0.936 & 2.829 & 11.20 & 1.57 &  9.6 &  842 &     & M\\
105  & 0.093 & 0.137 & 0.269 & 0.831 & 2.802 & 12.02 & 2.16 &  9.9 &  938 &     & M\\
 85  & 0.055 & 0.253 & 0.150 & 0.534 & 2.684 & 12.63 & 2.71 &  9.9 &  964 &-0.1 & M\\
218  & 0.116 & 0.078 & 0.192 & 0.959 & 2.858 & 11.59 & 1.65 &  9.9 &  973 &     & M\\
106  & 0.088 & 0.213 & 0.164 & 0.778 & 2.713 & 11.00 & 1.06 &  9.9 &  973 & 0.0 & M\\
 47  & 0.060 & 0.323 & 0.117 & 0.377 & 2.613 & 13.39 & 3.42 & 10.0 &  985 &-0.7 & M\\
 37  & 0.024 & 0.158 & 0.188 & 0.928 & 2.766 & 11.01 & 0.87 & 10.1 & 1065 &     & M\\
147  & 0.080 & 0.088 & 0.224 & 0.956 & 2.847 & 11.79 & 1.62 & 10.2 & 1080 &     & M\\
 70  & 0.089 & 0.209 & 0.166 & 0.858 & 2.713 & 10.54 & 0.31 & 10.2 & 1112 & 0.0 & M\\
 86  & 0.061 & 0.273 & 0.124 & 0.478 & 2.662 & 13.13 & 2.88 & 10.3 & 1120 &-0.5 & M\\
 26  & 0.067 & 0.130 & 0.206 & 0.972 & 2.795 & 11.12 & 0.84 & 10.3 & 1140 &     & M\\
 99  & 0.053 & 0.115 & 0.200 & 0.946 & 2.817 & 11.73 & 1.42 & 10.3 & 1154 &     & M\\
 18  & 0.113 & 0.123 & 0.178 & 0.943 & 2.806 & 11.63 & 1.18 & 10.5 & 1233 &     & M\\
115B & 0.261 &-0.041 & 0.247 & 0.737 & 2.759 & 10.43 &-0.22 & 10.6 & 1345 &     & M\\
 16  & 0.088 & 0.147 & 0.192 & 0.909 & 2.781 & 11.82 & 1.15 & 10.7 & 1357 &     & M\\
 54  & 0.052 & 0.214 & 0.152 & 0.747 & 2.713 & 12.13 & 1.43 & 10.7 & 1384 & -0.2 & M\\
113  & 0.082 & 0.151 & 0.168 & 0.974 & 2.767 & 11.12 & 0.37 & 10.8 & 1409 &  & M\\
 92  & 0.000 & 0.226 & 0.166 & 0.722 & 2.704 & 12.26 & 1.49 & 10.8 & 1423 & 0.0 & M\\
 46  & 0.040 & 0.289 & 0.128 & 0.525 & 2.645 & 12.79 & 2.01 & 10.8 & 1431 & -0.5 & M\\
 33  & 0.080 & 0.152 & 0.173 & 0.934 & 2.771 & 11.57 & 0.79 & 10.8 & 1431 &  & M\\
114  & 0.090 & 0.139 & 0.168 & 0.949 & 2.786 & 11.76 & 0.87 & 10.9 & 1506 &  & M\\
049B & 0.244 & -0.026 & 0.215 & 0.932 & 2.822 & 11.37 & 0.48 & 10.9 & 1513 &  & M\\
  5  & 0.015 & 0.153 & 0.189 & 0.947 & 2.770 & 11.73 & 0.78 & 11.0 & 1549 &  & M\\
  9  & 0.030 & 0.136 & 0.227 & 0.953 & 2.790 & 11.95 & 1.00 & 11.0 & 1551 &  & M\\
 62  & 0.008 & 0.216 & 0.165 & 0.726 & 2.714 & 12.81 & 1.72 & 11.1 & 1650 & 0.0 & M\\
139  & 0.080 & 0.278 & 0.106 & 0.553 & 2.646 & 12.85 & 1.57 & 11.3 & 1808 & -0.7 & M\\
 44  & 0.004 & 0.165 & 0.183 & 0.934 & 2.757 & 12.03 & 0.72 & 11.3 & 1824 &  & M\\
 53  & 0.023 & 0.155 & 0.198 & 0.893 & 2.774 & 12.66 & 1.31 & 11.4 & 1860 &  & M\\
 45  & 0.012 & 0.192 & 0.198 & 0.829 & 2.735 & 12.65 & 1.30 & 11.4 & 1863 &  & M\\
 78  & 0.004 & 0.266 & 0.168 & 0.566 & 2.676 & 13.65 & 2.27 & 11.4 & 1882 & 0.1 & M:\\
 39  & 0.028 & 0.184 & 0.165 & 0.955 & 2.728 & 11.41 &-0.03 & 11.4 & 1938 &  & NM\\
 41  & 0.000 & 0.176 & 0.195 & 0.956 & 2.741 & 12.01 & 0.31 & 11.7 & 2192 &  & NM\\
  8  & 0.039 & 0.317 & 0.154 & 0.490 & 2.630 & 13.94 & 2.20 & 11.7 & 2230 & -0.3 & NM\\
 56  & 0.013 & 0.234 & 0.153 & 0.784 & 2.691 & 12.11 & 0.36 & 11.8 & 2238 & -0.1 & NM\\
 36  & 0.000 & 0.344 & 0.180 & 0.561 & 2.614 & 13.22 & 0.98 & 12.2 & 2805 & -0.1 & NM\\
 42  & 0.000 & 0.231 & 0.184 & 0.829 & 2.698 & 12.57 & 0.22 & 12.4 & 2952 & 0.3 & NM\\
004B & 0.296 &-0.037 & 0.216 & 0.805 & 2.706 & 11.41 &-1.33 & 12.7 & 3532 &  & NM\\
 51  & 0.000 & 0.342 & 0.139 & 0.592 & 2.590 & 13.42 &-0.09 & 13.5 & 5033 & -0.6 & NM\\
 43  & 0.000 & 0.279 & 0.176 & 0.785 & 2.658 & 12.74 &-0.77 & 13.5 & 5041 & 0.1 & NM\\
040B & 0.161 &-0.050 & 0.195 & 0.601 & 2.673 & 12.38 &-1.76 & 14.1 & 6728 &  & NM\\
Mean value     & 0.074 &       &       &       &       &       &      & 10.42 & 1258 & -0.3\\
$\sigma$ & 0.057 &       &       &       &       &       &      & 0.61 & 339 & 0.3 & \\
\end{longtable}

%% file: Tbl8.tex
\setcounter{table}{7}
\tablecols{12}
\tabcaption{Cross ID and membership probabilities of
the open cluster NGC 6811}
\def\ColumnHeaders{
WBD & Luo & Sanders & Becker & Barkhatova & Kharchenko & Prob Sanders & Pkin & Pph & Psp & mbr\\ }
\begin{longtable}{rcccccclllll}
  \toprule
  \ColumnHeaders\\ \midrule
  \endfirsthead

  \tabcaptioncontinued
  \toprule
  \ColumnHeaders\\ \midrule
  \endhead

  \bottomrule
  \endfoot

112 &    & 177 & 51 & 1136 & & 0 & & & & nm\\
122 &    &     & & & & & & & & nm\\
107 &    & 183 & 48 & & & 96 & & & & nm\\
 31 &    & 102 & 15 & 1057 &  & 97 &  &  &  & m\\
489 & 6  & 205 & & 2080 & & 89 & & & & m\\
146 &    & 187 & 42 & 1143 & & 0 & & & & m\\
 68 &    & 110 & 86 & 1065 & 78 & 97 & 0.9272 & 1 & 1 & m\\
 34 &    & 121 & 22 & 1084 &  & 96 &  &  &  & m\\
105 &    & 185 & 47 & 1141 & & 25 & & & & m\\
 85 &    &  94 & 4 & 1041 &  & 95 &  &  &  & m\\
218 &    & 192 & & 1154 & & 92 & & & & m\\
106 &    & 172 & 49 & 1133 & & 97 & & & & m\\
 47 &    & 158 & 35 & 1121 &  & 96 &  &  &  & m\\
 37 & 2  & 136 & 26 & 1104 &  & 96 &  &  &  & m\\
147 &    & 189 & 43 & 1146 & & 97 & & & & m\\
 70 & 3  & 108 & 82 & 1063 & 76 & 96 & 0.9837 & 1 & 1 & m\\
 86 &    &  98 & 5 & 1050 &  & 89 &  &  &  & m\\
 26 &    &  86 & 10 & 1038 & 69 & 97 & 0.4193 & 1 & 1 & m\\
 99 &    & 159 & 29 & 1122 &  & 97 &  &  &  & m\\
 18 & 1  & 97 & 85 & 1049 & 73 & 97 & 0.0016 & 1 & 1 & m\\
115 &    & 149 & 65 & & & 95 & & & & m\\
 16 &    & 103 & 81 & 1059 &  & 90 &  &  &  & m\\
 54 &    & 147 & 68 & 1114 &  & 97 &  &  &  & m\\
113 & 5  & 166 & 60 & 1126 & 98 & 93 & 0.8128 & 1 & 1 & m\\
 92 &    & 115 & 20 & 1071 &  & 0 &  &  &  & m\\
 46 &    & 160 & 36 & 1124 &  & 96 &  &  &  & m\\
 33 & 13 & 113 & 19 & 1066 & 79 & 95 & 0.9765 & 0.9971 & 1 & m\\
114 &    & 146 & 66 & & & 96 & & & & m\\
 49 &    & 165 & 46 & 1125 & 97 & 93 & 0.7504 & 1 & 1 & m\\
  5 &    & 127 & 25 & 1092 & 83 & 93 & 0.9227 & 0.7357 & 1 & m\\
  9 & 16 & 134 & 71 & 1101 & 86 & 88 & 0.9565 & 1 & 1 & m\\
 62 &    & 123 & 77 & 1087 &  & 97 &  &  &  & m\\
139 &    & 131 & & 1096 & & 95 & & & & m\\
 44 & 11 & 157 & 33 & 1120 & 95 & 96 & 0.9137 & 1 & 1 & m\\
 53 & 10 & 135 & 69 & 1105 &  & 94 &  &  &  & m\\
 45 &    & 155 & 34 & 1118 & 94 & 97 & 0.0001 & 0.6244 & 1 & m\\
 78 &    &  72 & 11 & 1023 &  & 95 &  &  &  & m:\\
 39 & 4  & 144 & 30 & 1112 &  & 54 &  &  &  & nm\\
 41 &    & 154 & 28 & 1117 &  & 0 &  &  &  & nm\\
  8 &    &     &  & 1106 &  &  &  &  &  & nm\\
 56 &    & 138 & 67 & 1108 & 89 & 0 & 0.7463 & 0.9707 & 1 & nm\\
 36 &    & 132 &  & 1099 &  & 97 &  &  &  & nm\\
 42 & 14 & 143 & 31 & 1111 &  & 90 &  &  &  & nm\\
  4 & 12 & 119 & 23 & 1077 &  & 95 &  &  &  & nm\\
 51 & 15 & 137 & 70 & 1109 &  & 97 &  &  &  & nm\\
 43 &    & 142 & 32 & 1110 &  & 96 &  &  &  & nm\\
 40 &    &     &  &  &  &  &  &  &  & nm\\
  6 &    &     &  & 1093 &  &  &  &  &  & und\\
  7 &    &     &  & 1102 &  &  &  &  &  & und\\
 10 &    &     &  & 1085 &  &  &  &  &  & und\\
 12 &    &     &  & 1078 &  &  &  &  &  & und\\
 13 &    &     &  & 1082 &  &  &  &  &  & und\\
 14 &    & 114 & 80 & 1067 &  & 0 &  &  &  & und\\
 22 &    &  84 & 18 & 1029 &  & 0 &  &  &  & und\\
 23 & 18 &     & 17 & 1053 &  &  &  &  &  & und\\
 24 &    &  95 & 14 & 1047 & 72 & 97 & 0.8117 & 1 & 1 & und\\
 32 &    & 106 & 16 & 1061 &  & 97 &  &  &  & und\\
 35 &    & 128 &  & 1095 &  & 80 &  &  &  & und\\
 38 &    & 150 & 27 & 1115 &  & 0 &  &  &  & und\\
 57 &    & 129 & 72 & 1097 &  & 79 &  &  &  & und\\
 58 &    & 130 & 73 & 1098 &  & 71 &  &  &  & und\\
 63 &    &     &  &  &  &  &  &  &  & und\\
 64 &    &     &  &  &  &  &  &  &  & und\\
 65 &    & 120 & 78 & 1081 &  & 96 &  &  &  & und\\
 71 &    &  77 & 100 & 1026 & 66 & 97 & 0.5865 & 1 & 1 & und\\
 73 &    &  64 & 101 & 1016 & 61 & 0 & 0.0646 & 1 & 1 & und\\
 74 &    &  74 & 13 &  &  & 0 &  &  &  & und\\
 77 &    &  67 & 12 & 1018 &  & 97 &  &  &  & und\\
 79 &    &  85 & 1 & 1031 & 68 & 0 & 0.4112 & 1 & 1 & und\\
 82 &    &  89 & 3 & 1036 &  & 27 &  &  &  & und\\
 87 &    & 100 & 7 & 1054 &  & 57 &  &  &  & und\\
101 &    & 170 & 38 & 1131 & 87 & 97 & 0.8326 & 1 & 1 & und\\
102 &    & 179 & 40 & 1137 & & 0 & & & & und\\ 118 & &
145 & 64 &1113 & & 0 & & & & und\\
123 &    &     & & 1051 & & & & & & und\\
133 &    &  92 & 2 & 1039 & 71 & 94 & 0.8766 & 1 & 1 & und\\
178 &    &  76 & 99 & 1025 & 65 & 96 & 0.8519 & 0.9999 & 1 & und\\
491 & 7  & 209 & & & & 0 & & & & und\\
v17 &          & & & & & & & & & und\\
\end{longtable}